\documentclass[aps,twocolumn,superscriptaddress,nofootinbib]{revtex4}
\begin{document}

\title{Quantum entanglement, unitary braid representation\\
and Temperley-Lieb algebra}

\author{C.-L. Ho
}
 \affiliation{Department of Physics, and Center for Quantum
 Technologies\\
 National University of Singapore, 117543, Singapore}
\affiliation{Department of Physics, Tamkang University, Tamsui
251, Taiwan, R.O.C.\footnote{Permanent address}}

\author{A.I. Solomon}
\affiliation{Department of Physics and Astronomy, The Open
University, \\
Walton Hall, Milton Keynes MK7 6AA, UK} \affiliation{LPTMC,
Universit\'e Pierre et Marie Curie, 75252 Paris Cedex 05, France}

\author{C.-H. Oh}
\affiliation{Department of Physics, and Center for Quantum
 Technologies\\
 National University of Singapore, 117543, Singapore}

\date{Jun 3, 2010}

\begin{abstract}

Important developments in fault-tolerant quantum computation using
the braiding of anyons have placed the theory of braid groups at
the very foundation of topological quantum computing. Furthermore,
the realization by Kauffman and Lomonaco that a specific braiding
operator from the solution of the Yang-Baxter equation, namely the
Bell matrix, is universal implies that in principle all quantum
gates can be constructed from braiding operators together with
single qubit gates. In this paper we present a new class of
braiding operators from the Temperley-Lieb algebra that
generalizes the Bell matrix to multi-qubit systems, thus unifying
the Hadamard and  Bell matrices within the same framework.  Unlike
previous braiding operators, these new operators  generate {\it
directly}, from separable basis states,  important entangled
states such as the generalized Greenberger-Horne-Zeilinger states,
cluster-like states, and other states with varying degrees of
entanglement.

\end{abstract}


\maketitle

{\bf Introduction.}-- Recent developments in fault-tolerant
quantum computation using the braiding of anyons \cite{TQC}, have
stimulated interest in applying the theory of braid groups to the
fields of quantum information and quantum computation.  In this
respect, an interesting result is the realization that a specific
braiding operator is  a universal gate for quantum computing in
the presence of local unitary transformations \cite{Kauff04}. This
operator involves a unitary matrix $R$ that generates the four
maximally entangled Bell states from the standard basis of
separable states. This has led  to further investigation on the
possibility of generating other entangled states  by appropriate
braiding operators \cite{ZKG,CXG07,ZJG}.  In \cite{CXG07}, unitary
braiding operators were used to realize entanglement swapping and
generate the Greenberger-Horne-Zeilinger (GHZ) state \cite{GHZ},
as well as the linear cluster states \cite{cluster}.  Further
generalizations of the braiding operators to bipartite quantum
systems with states of arbitrary dimension, i.e., qudits, were
obtained by the approach of Yang-Baxterization \cite{ACDM,Xue}.

The GHZ state was not directly generated by the braiding operator
in \cite{CXG07}. The resulting state was transformed,   by use of
a local unitary transformation, to the GHZ state. We argue here
that this state does not, in fact, possess the same entanglement
properties as the GHZ state.  In this note we show how the Bell
states, the generalized GHZ states and some cluster-like states
may be generated {\em directly} from a braiding operator. We adopt
a different approach, based on the Temperley-Lieb algebra (TLA)
\cite{TL}, to obtain a class of unitary representations of the
braid group, and with it the required braiding operator. We first
obtain an explicit representation of the TLA, and then find the
braid group representation via the Jones representation
\cite{Jones85}.

\medskip

{\bf Braid group and quantum entanglement.}-- The $m$-stranded
braid group $B_m$ is generated by a set of elements
$\{b_1,b_2,\ldots,b_{m-1}\}$ with  defining relations:
\begin{eqnarray}
b_ib_j &=& b_jb_i,~~|i-j|>1;\nonumber\\
b_ib_{i+1}b_i &=& b_{i+1} b_i b_{i+1},~~1\leq i <m. \label{BGR}
\end{eqnarray}
Quantum computing requires that quantum gates be represented by
unitary operators.  Thus, for applications of the braid group in
quantum computation, one requires its unitary representations. For
an $m$-qubit system the usual $2^m\times2^m$ unitary
representation of $B_m$ employed in the literature is
\begin{eqnarray}
b_i= I \otimes\ldots \otimes I \otimes\ R \otimes I\otimes \ldots
\otimes I
\label{R} \; \; \; \; \; (i=1\ldots m-1)
\end{eqnarray}
where $I$ is the $2\times 2$ unit matrix and $R$ is a $4\times 4$
unitary matrix that acts on both the $i$-th and $(i+1)$-th qubits;
that is, occupying the $(i,i+1)$ position.  The first of the two
braid group relations in (\ref{BGR}) is automatically satisfied by
the form (\ref{R}). To fulfill the second relation, $R$ must
satisfy
\begin{eqnarray}
\left(R\otimes I\right)\left(I\otimes R\right)\left(R\otimes
I\right)=\left(I\otimes R\right)\left(R\otimes
I\right)\left(I\otimes R\right).\label{YBE}
\end{eqnarray}
This relation is sometimes called the (algebraic) Yang-Baxter
equation. One of the simplest solutions of (\ref{YBE}) that
produces entanglement of states is the matrix
\begin{eqnarray}
R= \frac{1}{\sqrt{2}}
 \left( \begin{array}{cccc}
  1 & 0 & 0 &-1 \\
  0 & 1 &-1 & 0 \\
  0 & 1 & 1 & 0 \\
  1 & 0 & 0 & 1
  \end{array}\right).
  \label{Bell-Mat}
\end{eqnarray}
When acting on the standard basis $\{|00\rangle, |01\rangle,
|10\rangle, |11\rangle\}$, $R$ generates the four maximally
entangled Bell states $(|00\rangle \pm |11\rangle)/\sqrt{2}$ and
$(|01\rangle \pm |10\rangle)/\sqrt{2}$. Here we adopt the
convention $|0\rangle=(1,0)^t$ and $|1\rangle=(0,1)^t$, where $t$
denotes the transpose. Following \cite{ZJG} and \cite{ACDM}, we
shall call $R$ the Bell matrix\footnote{Not to be confused with
the Bell matrix of combinatorial mathematics (after E.T. Bell).}.
In the presence of local unitary transformations, $R$ is a
universal gate \cite{Kauff04}.

The representation (\ref{R}) can also be used to generate
maximally entangled $n$-qubit states which are  equivalent, up to
local unitary transformation, to the GHZ states \cite{CXG07}. To
see this, let us take the $n=3$ qubit case, and consider the action
of $b_1b_2$ on the separable state $|000\rangle$:
\begin{eqnarray}
|\psi\rangle=b_1b_2|000\rangle=\frac{1}{2}\left(|000\rangle +
|011\rangle + |101\rangle + |110\rangle\right).\label{GHZ-like}
\end{eqnarray}
$|\psi\rangle$ is related to the GHZ state
$|GHZ\rangle=(|000\rangle + |111\rangle)/\sqrt{2}$ by a local
unitary transformation as
\begin{eqnarray}
|\psi\rangle=H\otimes H \otimes H
|GHZ\rangle,~~~H=\frac{1}{\sqrt{2}}\left(
\begin{array}{cc}
    1 & 1 \\
    1 & -1  \\
\end{array}\right)
\end{eqnarray}
where $H$ is the Hadamard matrix (or gate).

That the state $|\psi\rangle$ is said to be equivalent to the GHZ
state is based on the fact that local unitary transformations  do
not alter the degree of entanglement\footnote{A more precise statement is that
for bipartite states entanglement is preserved under LOCC (local operations and
classical communication).}. Nevertheless, it is evident
that they have very different entanglement properties.  For instance, after
making a measurement on any one of the three qubits, the other
two qubits of the GHZ state become separable, whereas those of
$|\psi\rangle$ are still in one of the maximally entangled Bell
states! It would be more desirable if one could generate the GHZ
states {\em directly} from the braiding operators without recourse to
any local unitary transformation.

A common feature of the Bell states and the GHZ states is that
they have the form of the superposition of a separable product
state $|a_1a_2\ldots a_k\cdots
a_n\rangle\equiv|a_1\rangle|a_2\rangle\cdots|a_n\rangle$ with its
conjugate state $|\bar{a}_1\bar{a}_2\ldots \bar{a}_k \cdots
\bar{a}_n\rangle$, which has all $a_k$'s changed from $0$ to $1$,
and $1$ to $0$, i.e., $\bar{a}_k=1,0$ if $a_k=0,1$, respectively.
Thus the state $|00\rangle$ is conjugate to $|11\rangle$,
$|001\rangle$ is conjugate to $|110\rangle$, etc.  As pointed out
after Eq.(\ref{Bell-Mat}), the Bell matrix  essentially
superimposes each two-qubit basis state on its conjugate, as does
the Hadamard matrix in the one-qubit case.

We wish to generalize the Hadamard and Bell matrices to higher
dimensions (i.e., to $n$ qubits), so that they generate
generalized GHZ states from separable states directly.  We want
these matrices to be representatives of certain braiding operators
of the braid group. Hence the main task is to find an appropriate
unitary representation of the braid group, and to determine the
correct combination of the braid generators that gives the
required matrix.  We find that a very simple way to achieve this
task is by means of the Jones representation of the braid group,
which we describe below.

\medskip

{\bf Unitary Jones representation of $B_3$.}-- In his construction
of certain polynomial invariants, the Jones polynomials, for knots
and links, Jones \cite{Jones85} provided a new representation of
the braid group based on what is essentially the TLA. The TLA,
more specifically denoted by $TL_m(d)$, is defined, for an integer
$m$ and a complex number $d$, to be the algebra generated by the
unit element $I$ and the elements $h_1,h_2,\ldots,h_{m-1}$
satisfying the relations
\begin{eqnarray}
h_ih_j&=&h_jh_i,~~|i-j|>1;\nonumber\\
h_ih_{i\pm 1}h_i &=& h_i,~~1\leq i <m, \label{TLA}\\
h_i^2 &=& dh_i.\nonumber
\end{eqnarray}

Given a TLA, the Jones representation of the braid group is
defined by (see eg., \cite{Kauff1})
\begin{eqnarray}
b_i=Ah_i + A^{-1} I,~~ b_i^{-1}=A^{-1}h_i + AI,\label{JonesRep}
\end{eqnarray}
where $A$ is a complex number given by $d=-A^2-A^{-2}$. It is
easily checked that the $b_i$'s so defined do satisfy the braid group
relation (\ref{BGR}).

In general the Jones representation is  not unitary.  However, it
is obvious from (\ref{JonesRep}) that if $A=e^{i\theta}$
($\theta\in [0,2\pi)$) and all the $h_i$'s are Hermitian
($h_i^\dagger =h_i$), then indeed the Jones representation is
unitary{\footnote{This representation is not faithful in that more
than one group element can be represented by the same matrix. It
is easily checked using the TLA and the binomial theorem that if
$m$ is the least integer such that $A^m=1$, then
$b_i^m=\left(\frac{(-1)^m-1}{d}\right) h_i + I$. Hence $b_i^m=I$
for $m$ even and $b_i^{2m}=I$ for $m$ odd. And so $b_i^k$ and
$b_i^l$ have the same matrix representation if $k$ and $l$ differ
by a multiple of $m$ ($m$ even) or $2m$ ($m$ odd). Similarly, the
commonly used representation (\ref{R}) with $R$ given by
(\ref{Bell-Mat}) is also not a faithful representation, since
$R^8=I$ implies $b_i^8=I$. However, one can obtain a faithful
representation $\hat{b}_i$ by defining $\hat{b}_i\equiv e^{\theta}
b_i$, where $\theta/\pi$ is irrational but otherwise arbitrary. }.

Based on this fact, in what follows we shall provide a class of
unitary representation of the 3-stranded braid group $B_3$, and
show that a subclass of it gives nonlocal unitary transformations
that generate conjugate-state entanglements from separable basis
states.

For $A=e^{i\theta}$, $d=-2\cos 2\,\theta$ is real.  A simple
unitary representation of $B_3$ is given by the Jones
representation with TLA elements $h_i=d E_i$ ($i=1,2$), where
\begin{eqnarray}
E_1 =\left( \begin{array}{cc} 1 & 0\\0 & 0
\end{array}\right),~~~
E_2 = \left( \begin{array}{cc} a^2 & e^{-i\phi} ab\\e^{i\phi} ab &
b^2
\end{array}\right),~~a^2+b^2=1.\label{E12}
\end{eqnarray}
Here $\phi$ is a phase angle. The $E_i$'s satisfy
\begin{eqnarray}
E_i^2 &=& E_i,\nonumber\\
E_1E_2E_1 &=& a^2 E_1,\label{E-alg}\\
E_2E_1E_2 &=& a^2 E_2.\nonumber
\end{eqnarray}
With $a^2=d^{-2}$, $h_i$'s as constructed from $E_i$'s satisfy the
TLA. Now as $d$ and $a$ are real, in order that $h_i$'s be
Hermitian, we must have $b^2=1-1/d^2 \geq 0$. This implies
$d^2\geq 1$, and hence $\theta ({\rm mod}\ 2\pi)$ is restricted to
be in the range $|\theta|\leq \pi/6$ or $|\theta-\pi|\leq \pi/6$.
We shall assume $\theta$ to be in these domains below.  The
special case of this representation with $\phi=0$ was employed
previously in exploring the relation between quantum computing and
the Jones polynomials \cite{Kauff1} (see also \cite{Kauff2}).

A very simple way to generalize the above representation of TLA to
higher dimensions is as follows.  Let
\begin{eqnarray}
e_1= \left( \begin{array}{cc} 1 & 0\\0 & 0
\end{array}\right),\,e_2=\left( \begin{array}{cc} a^2 & 0\\ 0 & b^2
\end{array}\right),\,e_3=\left( \begin{array}{cc}  0 & e^{-i\phi} \\ e^{i\phi} &
0\end{array}\right).
\end{eqnarray}
Define
\begin{eqnarray}
E_1^{(n,k)}&\equiv& \otimes_{j=1}^{k-1}I \otimes e_1
\otimes_{j=k+1}^n I,\label{E-1}\\
E_2^{(n,k)} &\equiv& \otimes_{j=1}^{k-1}I \otimes e_2
\otimes_{j=k+1}^n I\nonumber\\
&+& ab \otimes_{j=1}^{k-1} s_j \otimes e_3 \otimes_{j=k+1}^n
s_j,\label{E-2}
\end{eqnarray}
where $\otimes_{j=1}^{m} s_j =s_1\otimes s_2 \otimes \cdots
\otimes s_m$. Here $s_j$ is any Hermitian operator satisfying
$s_j^2=1$. For example, $s_j$ can be $I$, any one of the Pauli
matrices $\sigma_m (m=1,2,3)$, or the Hadamard matrix $H$. The
integer $n$ is the number of $2\times 2$ matrices in the tensor
products, and $k$ indicates the position of $e_1,~e_2$ and $e_3$.
The $E_i^{(n,k)}$'s are $2^n\times 2^n$ matrices, and they reduce
to (\ref{E12}) in the case $n=k=1$.  One can easily check that
$E_i^{(n,k)}$'s satisfy (\ref{E-alg}).  Hence, the operators
$h^{(n,k)}_i=d E_i^{(n,k)}$ form a $2^n\times 2^n$ matrix
realization of $TL_3 (d)$\footnote{See \cite{Xue} for an
$n^2\times n^2$ matrix realization of the TLA. The braiding
operator (called the Yang-Baxter matrix in these works) was
obtained there through a Yang-Baxterization process. This latter
process was also employed in \cite{ACDM}, but not related to TLA,
to generalize the Bell matrix to $(2n)^2\times (2n)^2$ braid
matrices.}.  A unitary braid group representation is then obtained
from the $h_i$'s by the Jones representation.

Our new unitary braid representation generalizes the $2\times 2$
matrices of (\ref{E12}) to $2^n\times 2^n$ matrices of (\ref{E-2})
within the TLA $TL_3 (d)$. Other routes of generalization are
possible. For instance, in \cite{Kauff3} the $2\times 2$
representation of $TL_3(d)$ were generalized to higher dimensional
matrices for $TL_m (d)$ with $m>3$, where the dimension of
representation varies with the number of strands $m$ according to
the Fibonacci numbers, or with the number of independent
bit-strings of certain path model proposed in \cite{AJL}.

{\bf Generalized GHZ states.}-- From now on we will be mainly
concerned with the unitary braiding transformation representing
the action of the braid $b_1b_2$. This braiding operator can be
evaluated to be
\begin{eqnarray}
&& b_1^{(n,k)}b_2^{(n,k)}\nonumber\\
 &=& \otimes_{j=1}^{k-1}I
\otimes
 \left( \begin{array}{cc}
  d a^2 & 0\\0 & d b^2 + A^{-2}
\end{array}\right)
\otimes_{j=k+1}^n I\label{b1b2}\\
&+& \otimes_{j=1}^{k-1} s_j \otimes
 \left( \begin{array}{cc}
 0 & -e^{-i\phi}A^4d ab \\ e^{i\phi}dab & 0
\end{array}\right) \otimes_{j=k+1}^n
s_j.\nonumber
\end{eqnarray}
Its action on the separable $n$-qubit states $|a_1a_2\ldots
a_{k-1} 0 a_{k+1}\cdots a_n\rangle$ and $|a_1a_2\ldots a_{k-1} 1
a_{k+1}\cdots a_n\rangle$ ($a_j=0,1,~j=1,2,\ldots,k-1,k+1,\ldots
n$) is given by
\begin{eqnarray}
&&b_1^{(n,k)}b_2^{(n,k)}|a_1a_2\ldots a_{k-1} 0 a_{k+1}\cdots
a_n\rangle\nonumber\\
&=&(da^2)|a_1a_2\ldots a_{k-1} 0 a_{k+1}\cdots
a_n\rangle\nonumber\\
 &+& (e^{i\phi}dab)|\tilde{a}_1\tilde{a}_2\ldots \tilde{a}_{k-1} 1 \tilde{a}_{k+1}\cdots
\tilde{a}_n\rangle,\label{act-1}
\end{eqnarray}
and
\begin{eqnarray}
&&b_1^{(n,k)}b_2^{(n,k)}|a_1a_2\ldots a_{k-1} 1 a_{k+1}\cdots
a_n\rangle\nonumber\\
&=&(d b^2 + A^{-2})|a_1a_2\ldots a_{k-1} 1 a_{k+1}\cdots
a_n\rangle\nonumber\\
& + & (-e^{-i\phi}A^4d ab)|\tilde{a}_1\tilde{a}_2\ldots
\tilde{a}_{k-1} 0 \tilde{a}_{k+1}\cdots
\tilde{a}_n\rangle,\label{act-2}
\end{eqnarray}
where $|\tilde{a}_j\rangle\equiv s_j |a_j\rangle$
($j=1,\ldots,k-1,k+1,\ldots,n$). Thus under the action of
$b_1^{(n,k)}b_2^{(n,k)}$, the separable $n$-qubit state
$|a_1a_2\ldots  a_k\cdots a_n\rangle$ is superimposed on the
state $|\tilde{a}_1\tilde{a}_2\ldots \tilde{a}_k \cdots
\tilde{a}_n\rangle$ in either the form (\ref{act-1}) or
(\ref{act-2}), depending on whether the $k$-th qubit $|a_k\rangle$
is $|0\rangle$ or $|1\rangle$.  The states in (\ref{act-1}) and
(\ref{act-2}) are normalized, as $ (da^2)^2 + |e^{i\phi}dab|^2 =1,
$ and $ |d b^2 + A^{-2}|^2 + |-e^{-i\phi}A^4d ab|^2=1, $ which can
be easily checked. Depending on the choice of the set of $s_j$'s,
the resulting state (\ref{act-1}) or (\ref{act-2}) will have
varying degrees of entanglement. In particular, if all $s_j=I$,
then the resulting state is separable, and
$b_1^{(n,k)}b_2^{(n,k)}$ is simply a local unitary transformation.

We now consider a subclass of the representation obtained by
setting $\phi=0$ in (\ref{E-2})  (i.e., $e_3=\sigma_1$), $s_j=I$
for $j<k$, and $s_j=\sigma_1$ for $j>k$ . In this case,
$|\tilde{a}_j\rangle=|a_j\rangle$ for $j<k$ and
$|\tilde{a}_j\rangle=\sigma_1 |a_j\rangle=|\bar{a}_j\rangle$ for
$j>k$. Hence, under the action of $B(n,k)\equiv
b_1^{(n,k)}b_2^{(n,k)}$ (with the above-mentioned choice of the
$s_j$'s in $b_i^{(n,k)}$ understood), the separable $n$-qubit
state $|a_1a_2\ldots a_{k-1}a_ka_{k+1}\cdots a_n\rangle$ is
superimposed on the state $|a_1a_2\ldots a_{k-1}\bar{a}_k
\bar{a}_{k+1} \cdots \bar{a}_n\rangle$ in either the form
(\ref{act-1}) or (\ref{act-2}) (with the appropriate change in the
$\tilde{a}_j$), depending on whether the $k$-th qubit
$|a_k\rangle$ is $|0\rangle$ or $|1\rangle$. The resulting states
are separable in the first $(k-1)$ qubits, but entangled in the
other $(n-k+1)$ qubits.   In particular, for $k=1$, the operator
$B(n,1)$ entangles the state $|a_1a_2\ldots a_k\cdots a_n\rangle$
with its conjugate state $|\bar{a}_1\bar{a}_2\ldots \bar{a}_k
\cdots \bar{a}_n\rangle$, thus giving the generalized GHZ states.
We see that these states can indeed be obtained from separable
basis states by the braiding operator.

We now give a few examples of the braiding operator $B(n,1)$ for
$k=1$ and $n=1, 2,3$.  From now on we choose $\theta=\pi/8$. This
gives $d=-\sqrt{2}$, and $a,b=\pm1/\sqrt{2}$.  Without loss of
generality, we take $a=b=1/\sqrt{2}$.  The four matrix elements in
(\ref{b1b2}) are $da^2=dab=-1/\sqrt{2}$ and
$db^2+A^{-2}=A^4dab=-i/\sqrt{2}$.  Explicitly, $B(n,1)$ has the
form
\begin{eqnarray}
B(n,1) &=& \left( \begin{array}{cc}
  -\frac{1}{\sqrt{2}} & 0\\0 & -\frac{i}{\sqrt{2}}
\end{array}\right)
\otimes_{j=2}^n I\label{B}\\
&+&
 \left( \begin{array}{cc}
 0 & \frac{i}{\sqrt{2}} \\ -\frac{1}{\sqrt{2}} & 0
\end{array}\right) \otimes_{j=2}^n
\sigma_1.\nonumber
\end{eqnarray}

For $n=1$, $B(1,1)=-\frac{1}{\sqrt{2}}\left(
\begin{array}{cc}
    1 &-i  \\
    1 & i  \\
    \end{array}\right)$ is, up to global phases, equivalent to the Hadamard
gate. For $n=2$:
\begin{eqnarray}
B(2,1) = -\frac{1}{\sqrt{2}}
 \left( \begin{array}{cccc}
  1 & 0 & 0 &-i \\
  0 & 1 &-i & 0 \\
  0 & 1 & i & 0 \\
  1 & 0 & 0 & i
  \end{array}\right).\label{Bell-equiv}
\end{eqnarray}
This is equivalent to the Bell matrix up to global phases, and it
gives all  four Bell states from the separable standard basis.
For example, when acting on the states $|00\rangle$ and
$|10\rangle$, it gives $-(|00\rangle + |11\rangle)/\sqrt{2}$ and
$-i(|10\rangle - |01\rangle)/\sqrt{2}$, respectively.

Note, however, the difference between the appearance of this
matrix in our approach, and the Bell matrix $R$ in
(\ref{Bell-Mat}). There the Bell matrix $R$ is the solution of the
algebraic Yang-Baxter equation (\ref{YBE}), and is the basic
building block of the braid generators $b_i$ in (\ref{R}).  In our
approach the matrix (\ref{Bell-equiv}) is obtained from the
product of the matrices representing the braid generators $b_1$
and $b_2$, i.e., it represents the braid $b_1b_2$.  In a sense, we
have factorized $R$.

It was mentioned in the Introduction that the main impetus to
using braid group representations in quantum computing is that the
Bell matrix is a universal gate \cite{Kauff04}.  Since $B(2,1)$ is
equivalent to $R$ in generating the Bell states, it should also be
a universal gate. To prove that, it suffices to show, following
\cite{Kauff04}, that the universal CNOT gate can be generated from
$B(2,1)$ and local unitary transformations.  This is indeed the
case, as we have ${\rm CNOT}=(\alpha\otimes
\beta)B(2,1)(\gamma\otimes \delta)$, where
\begin{eqnarray}
\alpha&=&\frac{1}{\sqrt{2}}\left(
\begin{array}{cc}
    1 &  i  \\
    1 & -i  \\
    \end{array}\right),~~~
\beta=\frac{1}{\sqrt{2}}\left(
\begin{array}{cc}
    1 &-i  \\
    i & -1  \\
    \end{array}\right),\nonumber\\
\gamma&=&\frac{1}{\sqrt{2}}\left(
\begin{array}{cc}
    -1 &  i  \\
     1 &  i  \\
    \end{array}\right),~~~
\delta=\left(
\begin{array}{cc}
    1 & 0  \\
    0 & -1  \\
    \end{array}\right).
\end{eqnarray}

Generalized GHZ states for larger $n$ can be obtained accordingly.

{\bf Cluster-like states.}-- As mentioned in the Introduction, the
GHZ state is rather fragile in its entanglement, as it becomes
separable after one of its qubits is measured. Multi-qubit systems
which possess more robust entanglement can in fact be generated
using $B(n,k)$. As an example, we consider the result of applying
a braiding operator $B(n,k)$ on a generalized GHZ state
$|\phi\rangle$ generated from $|00\ldots 00\rangle$ with the
braiding operator $B^{-1}(n,1)=B^\dagger (n,1)$.  We have
$|\Phi\rangle = B^{-1}(n,1)|00\ldots 00\rangle =(|00\ldots
00\rangle +i |11\ldots 11\rangle)/\sqrt{2}$. Upon applying
$B(n,k)$ to $|\phi\rangle$, we get
 \begin{eqnarray}
B(n,k)|\Phi\rangle = & \frac{1}{2}&\left(|00\cdots
00\rangle_{k-1}|00\cdots 00\rangle_{n-k+1}\right. \nonumber\\
& +&  \left. \, |00\cdots
00\rangle_{k-1}|11\cdots 11\rangle_{n-k+1}\right.\nonumber\\
&+& \left.  \,|11\cdots 11\rangle_{k-1}|00\cdots 00\rangle_{n-k+1}\right.
 \label{cluster}\\
 &-& \left.\, |11\cdots 11\rangle_{k-1}|11\cdots
11\rangle_{n-k+1}\right).\nonumber
\end{eqnarray}
Here $|00\cdots 00\rangle_{k-1}\equiv |0\rangle_1
|0\rangle_2\cdots |0\rangle_{k-1}$, $|00\cdots
00\rangle_{n-k+1}\equiv |0\rangle_k |0\rangle_{k+1}\cdots
|0\rangle_n$, etc.  This state is an entangled state for $n\geq 2$
and $k>1$. Unlike the GHZ states, when it loses one of its qubits,
the remaining state is still partially entangled when $n>2$.  For
$n=4$ and $k=3$, the state (\ref{cluster}) is just the $4$-qubit
linear cluster state given in \cite{cluster}.

By acting with $B(n,k)B^{-1}(n,1)$ on any one of the $2^n$
separable basis state $|a_1 a_2\cdots a_n\rangle$, one can in fact
generate $2^n$ orthogonal cluster-like states similar to those of
(\ref{cluster}) .

{\bf Summary.}-- In summary, we have obtained a new class of
unitary representation of the three-stranded braid group by the
Jones representation. The construction is based on a new matrix
realization of the Temperley-Lieb algebra. A subclass of the
representation provides a braiding operator that can superimpose
states on their conjugate states, thus giving the generalized GHZ
states. This braiding operator becomes the Hadamard matrix and the
Bell matrix in the one-qubit and two-qubit case, respectively.
Certain cluster-like states with robust entanglement can also be
generated from separable basis states with two such braiding
operators.

\begin{acknowledgments}

This work is supported in part by Singapore's A*STAR grant WBS
(Project Account No.) R-144-000-189-305, and in part by the
National Science Council (NSC) of the R.O.C. under Grant No. NSC
96-2112-M-032-007-MY3. CLH and AIS would like to thank the Centre
for Quantum Technologies and the Department of Physics at the
National University of Singapore for their hospitality.

\end{acknowledgments}

\end{document}